\documentclass[preprint,showpacs,preprintnumbers,amsmath,amssymb,superscriptaddress]{revtex4}

\usepackage{graphicx}
\usepackage{dcolumn}
\usepackage{bm}
\usepackage{hyperref}

\begin{document}

\title{Probing the dark exciton states of a single quantum dot using photocurrent spectroscopy in magnetic fields}

\author{Kai Peng}
\author{Shiyao Wu}
\affiliation{Beijing National Laboratory for Condensed Matter Physics, Institute of Physics, Chinese Academy of Science, Beijing 100190, China}
\affiliation{University of Chinese Academy of Sciences, Beijing 100049, China}
\author{Jing Tang}
\affiliation{Beijing National Laboratory for Condensed Matter Physics, Institute of Physics, Chinese Academy of Science, Beijing 100190, China}
\author{Feilong Song}
\author{Chenjiang Qian}
\author{Sibai Sun}
\author{Shan Xiao}
\affiliation{Beijing National Laboratory for Condensed Matter Physics, Institute of Physics, Chinese Academy of Science, Beijing 100190, China}
\affiliation{University of Chinese Academy of Sciences, Beijing 100049, China}
\author{Meng Wang}
\affiliation{Beijing National Laboratory for Condensed Matter Physics, Institute of Physics, Chinese Academy of Science, Beijing 100190, China}
\author{Ali Hassan}
\affiliation{Beijing National Laboratory for Condensed Matter Physics, Institute of Physics, Chinese Academy of Science, Beijing 100190, China}
\affiliation{University of Chinese Academy of Sciences, Beijing 100049, China}
\author{David A. Williams}
\affiliation{Hitachi Cambridge Laboratory, Cavendish Laboratory, Cambridge CB3 0HE, UK}
\author{Xiulai Xu}%
\email{xlxu@iphy.ac.cn}
\affiliation{Beijing National Laboratory for Condensed Matter Physics, Institute of Physics, Chinese Academy of Science, Beijing 100190, China}
\affiliation{University of Chinese Academy of Sciences, Beijing 100049, China}

\date{\today}

\begin{abstract}
We report on high-resolution photoluminescence (PL) and photocurrent (PC) spectroscopies of a single self-assembled InAs/GaAs quantum dot (QD) embedded in an n-i-Schottky device with an applied magnetic field in Faraday and Voigt geometries. The single-QD PC spectrum of neutral exciton (X$^0$) is obtained by sweeping the bias-dependent X$^0$ transition energy to achieve resonance with a fixed narrow-bandwidth laser through quantum-confined Stark effect. With a magnetic field applied in Faraday geometry, the diamagnetic effect and the Zeeman splitting of X$^0$ are observed both in PL and PC spectra. When the magnetic field is applied in Voigt geometry, the mixture of bright and dark states results in an observation of dark exciton states, which are confirmed by the polarization-resolved PL and PC spectra.
\end{abstract}

\pacs{78.67.Hc, 78.55.Cr,71.35.Ji}
\maketitle


\section{\label{sec:level1}Introduction}
Due to the three-dimensional confinement, semiconductor quantum dots (QDs) have atom-like discrete energy levels, which makes QDs a good candidate of quantum information processing \cite{loss1998quantum,imamog1999quantum}.
In-plane magnetic field has always been used to construct the energy level structure as a platform to manipulate the single spins in QDs \cite{xu2007fast,Berezovsky349,press2008,brunner2009coherent,Kim2010Fast,godden2012coherent,warburton2013single} by mixing bright and dark exciton states \cite{bayer2002fine,poem2010accessing,schwartz2015deterministic}. Additionally, an applied magnetic field on QDs lifts the degeneracy of electron and hole by Zeeman splitting, and shrinks the carrier wavefunctions by diamagnetic effect \cite{Tsai2008Diamagnetic,Cao2016}. By embedding QDs in a diode structure, photocurrent (PC) spectroscopy, as an effective way to detect the energy level and carriers¡¯ information in a single QD, has been used to observe Rabi oscillations of a two-level system in single QDs \cite{zrenner2002coherent,stufler2006two,Ramsay2008Fast,takagi2008coherently}, which can be utilized to build QD-based spin qubit. However, the finite dephasing time of the carriers in QDs makes the qubit to be manipulated by ultrashort laser pulses, which limits the resolution because of the wide spectral linewidth. In contrast, continuous-wave (CW) laser owns ultra-narrow-bandwidth that allows high-resolution measurement \cite{mar2013high,mar2017precise} and high-fidelity initialization of the QD spin qubit \cite{mar2014ultrafast} through PC spectroscopy.

In this letter, we present a systematic study of PL and PC spectroscopy of a single InAs QD embedded in the intrinsic region of an n-i-Schottky photodiode based on a two-dimensional electron gas (2DEG) with applying a magnetic field in Faraday and Voigt configurations. The high-resolution PC spectra of neutral exciton (X$^0$) are measured by sweeping the X$^0$ transition energy through quantum-confined Stark effect (QCSE) to achieve the resonant excitation with a fixed CW laser. With a vertical magnetic field (Faraday geometry) applied, the diamagnetic effect and the Zeeman splitting are observed through both PL and PC spectroscopies. When a magnetic field is applied in the QD plane (Voigt geometry), weak dark excitons states of neutral exciton and biexction (XX) are observed in PL spectra. High-resolution PC spectra of X$^0$ in Voigt geometry show the similar behavior, where the dark exciton states are confirmed with polarization-resolved PC spectroscopy.

\section{\label{sec:level1}EXPERIMENTAL DETAILS}
The n-i-Schottky device was designed for both PL and PC measurement of single QDs. A distributed Bragg reflector (DBR) of 13-pairs of Al$_{0.94}$Ga$_{0.06}$As/GaAs (67/71 nm) was grown below the structure, followed by a 200 nm intrinsic GaAs buffer layer. A Si $\delta$-doped GaAs layer with a doping density Nd = 5$\times$10$^{12}$ cm$^{-2}$ forms a 2DEG, which connects the AuGeNi alloy ohmic contact. Then a single layer of InAs QDs with a density less than 1$\times$10$^{9}$ cm$^{-2}$ were grown along the [001] crystallographic direction embedded in a 250-nm-thick intrinsic GaAs layer with a location of 50 nm above the 2DEG. In the active regime, a 10 nm semitransparent Ti was evaporated on the surface to form Schottky contact, then Al mask with apertures of about 1-3 $\mu$m were patterned to isolate single QDs. Finally, Cr/Au bond pads were fabricated on the Schottky and ohmic contact to bond electrodes. The device structure allows to apply vertical electrical field on the QDs with the field strength of \emph{F}=(\emph{V}$_i$-\emph{V}$_b$)/\emph{d}, where \emph{V}$_i$, \emph{V}$_b$ and \emph{d} are built-in potential, applied bias voltage and distance between the Schottky contact and 2DEG, respectively. For the device measured in this work in particular, the build-in potential \emph{V}$_i$ was measured as 0.8 V. More details can also be found in Ref. \cite{tang2014charge}.

The device was placed on a xyz piezoelectric stage in the helium gas exchange cryostat at 4.2 K equipped with superconducting magnets with magnetic fields up to 9 T in Faraday geometry and 4 T in Voigt geometry. To perform micro-PL measurement, a confocal microscopy was furnished with a large numerical aperture of NA=0.83 microscope objective. Non-resonant excitation was performed by using a 650 nm semiconductor laser for PL measurement, and the resonant pumping was achieved by a tunable narrow-bandwidth ($\sim$1 MHz) external cavity diode laser (ECDL) in Littrow configuration. The PL of QDs was collected and dispersed through a 0.55 m spectrometer, and the signal was detected with a liquid nitrogen cooled charge coupled device camera (CCD) with a spectra resolution of about 60 $\mu$eV. Half-wave plates, polarizers and quarter-wave plates were added in the collection or the pumping part to perform the polarization-depend PL and PC measurement. A source-measurement unit (SMU) with a high current resolution (10 fA) was used to apply bias voltage across the Schottky diode and measure the current.

\begin{figure}
\includegraphics[scale=0.9]{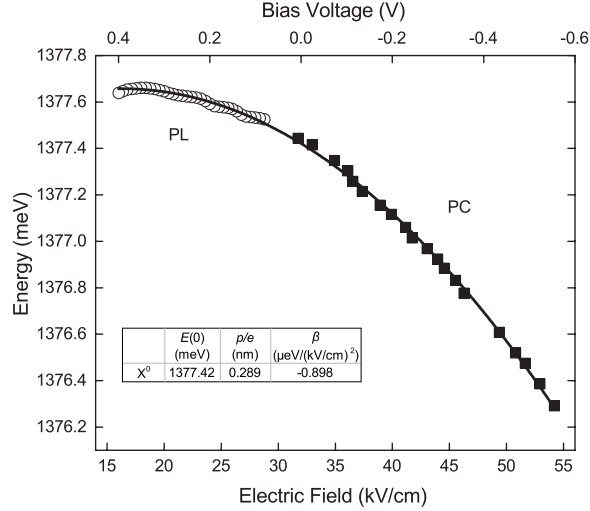}
\caption{\label{fig:1}Stark shift of X$^0$ as a function of electric field and bias voltage obtained from the single QD PL (circle) and PC (solid square) spectra. The solid line represents a quadratic fit by QCSE, which provides the correlation between bias voltage (electric field) and X$^0$ transition energy. Inset: The fitted parameters for \emph{E}$_0$, \emph{p}, and $\beta$ are given in the table.
}
\end{figure}

\section{\label{sec:level1}Results and discussion}

Before carrying out the PC measurements, bias-dependent micro-PL spectroscopy was performed on single QD. It provides the transition energies of different charged excitons states and the bias voltage range for PC regime. The X$^0$ and the XX transition energy were determined by power- and polarization-dependent PL measurement, and the fine structure splitting (FSS) is about 18 $\mu$eV at 0.4 V. The applied electric field tunes the band structure of QDs, which induces quantum-confined Stark effect (QCSE). The energy change of X$^0$ can be described as  \emph{E(F)} = \emph{E}(0) +\emph{pF} + $\beta$\emph{F}$^2$, where \emph{E}(0) is the transition energy without external applied field, \emph{p} is the permanent dipole moment and $\beta$ is the polarizability of electron-hole wavefunctions. To achieve resonant excitation, the energy of the ECDL is tuned a little lower than the X$^0$ energy extracted from PL spectra. Then the PC signal of X$^0$ is obtained by sweeping the energy level of X$^0$ via QSCE to be resonant with the fixed laser energy. The corresponding central voltage is obtained by fitting the extracted PC spectrum of X$^0$ with a Lorentzian curve. By tuning the pumping laser wavelength, a series of PC spectra are obtained. Fig.~\ref{fig:1} shows the Stark shift of X$^0$ in a single QD for both PL and PC regimes. With the fitted results in the inset, the correlation of electric field (bias voltage) applied on the QD and the energy level position of X$^0$ can be determined precisely.

\begin{figure}[b]
\includegraphics[scale=0.8]{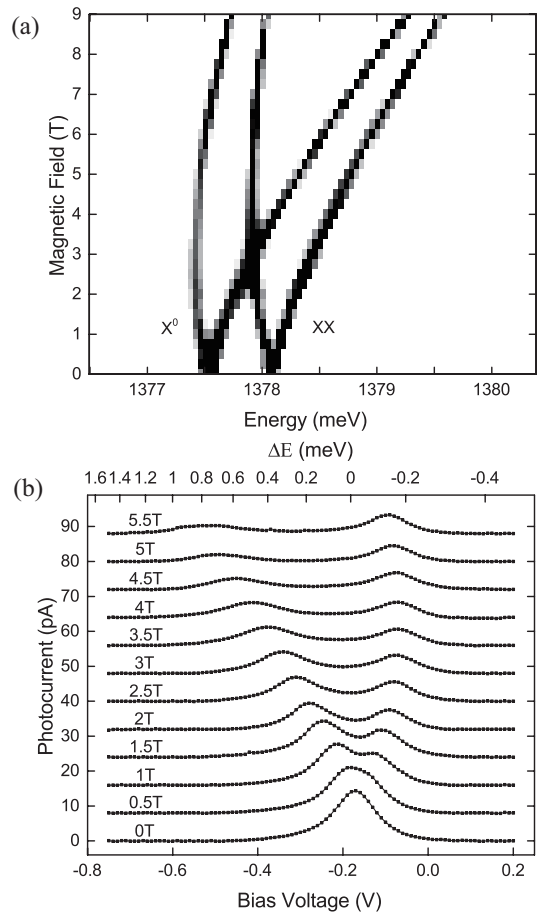}
\caption{\label{fig:2}(a) PL spectra of X$^0$ and XX with a magnetic field from 0 T to 9 T in Faraday geometry at 0.4 V. (b) PC spectra of X$^0$ with a magnetic field from 0 T to 5.5 T in Faraday geometry with pumping laser energy as \emph{E}$_{Laser}$=1377.21 meV. The top x-axis represents the energy shift comparing with the X$^0$ transition energy at zero magnetic field, which has been calculated according to QCSE. The spectra are shifted for clarity.
}
\end{figure}

\begin{figure}[b]
\includegraphics[scale=0.8]{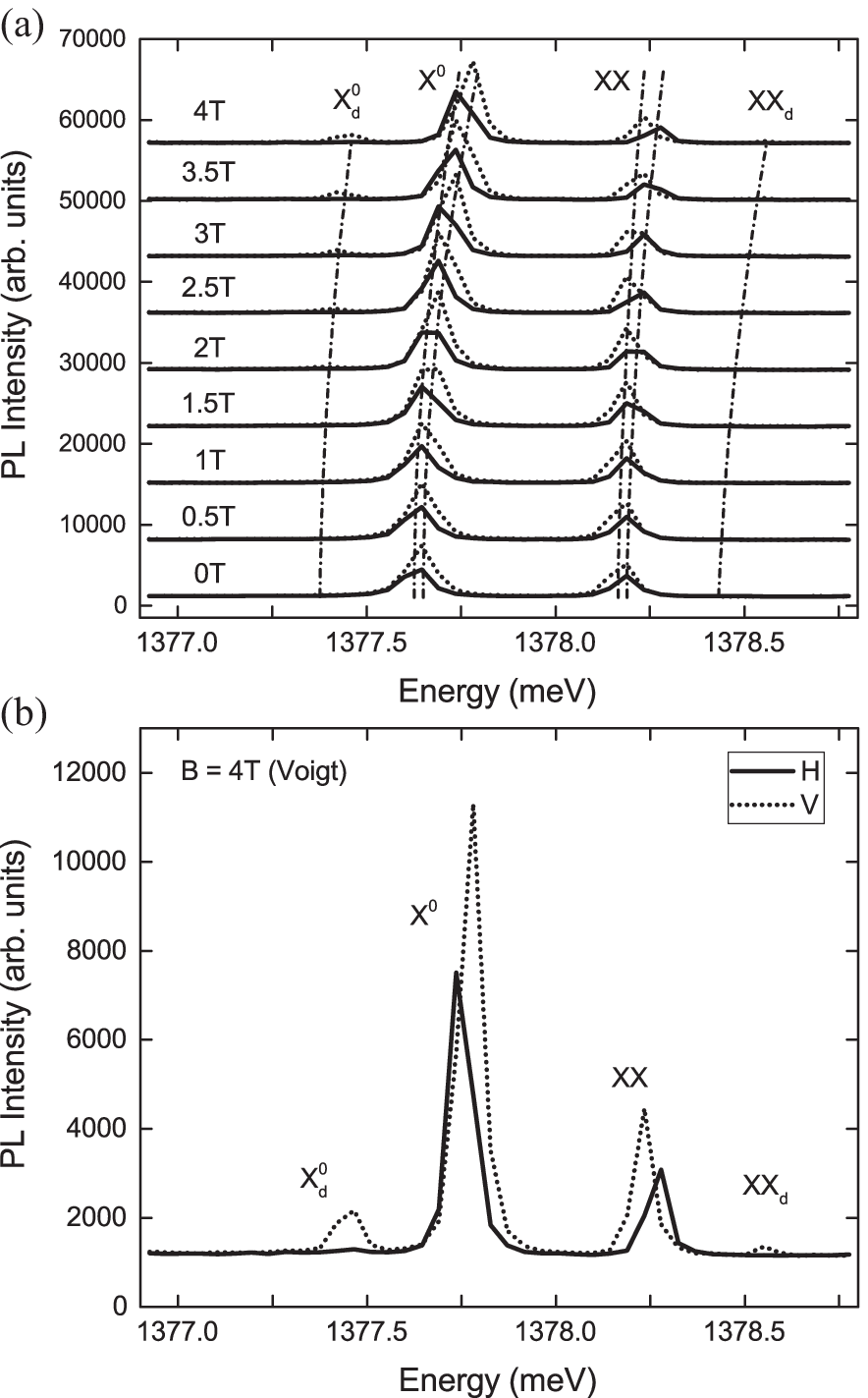}
\caption{\label{fig:3} (a) Polarization-resolved PL spectra of X$^0$ and XX with magnetic field from 0 T to 4 T in Voigt geometry at 0.4 V. The solid lines and dotted lines represent the H- and V-polarized PL spectra, respectively. The dash-dot lines gives the quadratic fitting of the PL peaks. The spectra are shifted for clarity. (b) Enlarged PL spectra at 4 T. The dark exciton emissions X$^0_d$ and XX$_d$ of both X$^0$ and XX can be clearly observed but the intensities are much weaker than bright exciton emission.
}
\end{figure}

When a magnetic field is applied along the growth direction, namely in Faraday geometry, the energy levels of X$^0$ and XX of a single QD split into two branches and moved toward high energy part, as shown by the PL spectra in Fig.~\ref{fig:2}(a). The energy splitting is caused by the Zeeman effect, which lifts off spin degeneracy of carriers. The excitons with different angular momenta of M=$\pm$1 combine and emit left and right circular polarized photons. Moving towards high energy is due to the diamagnetic effect, which is attributed to the magnetic field induced confinement of carrier wavefunctions in the QDs \cite{Cao2016}. It can be described as a quadratic relationship approximately. With the consideration of both Zeeman splitting and diamagnetic effect, the energy shift for X$^0$ can be described as: $\Delta$\emph{E}= $\gamma$\emph{B}$^2$$\pm$\emph{g}$_{ex}$$\mu$$_B$\emph{B}, where $\gamma$ is the diamagnetic coefficient, \emph{g$_{ex}$} is the g-factor of the exciton, and $\mu$$_B$ is the Bohr magneton.

Both effects also occur on PC spectrum with resonant pumping with a laser energy of \emph{E}$_{Laser}$=1377.21 meV, as shown in Fig.~\ref{fig:2}(b).
Here, in order to observe both branches, the linear polarized light is chosen to pump the X$^0$ resonantly. Because of the fixed energy of pumping laser, it is worth noting that, the energy level of X$^0$ are tuned lower to match the laser energy with an additional energy shift induced by the diamagnetic effect. As a result, the peaks of the PC spectra move to the negative bias voltage (high electric field) direction at high magnetic field. In Fig. ~\ref{fig:2}(b), the top x-axis represents the energy shift. Here the relationship is quadratic according to QCSE and the zero represents the X$^0$ transition energy at zero magnetic field. From these data, the absolute value of \emph{g$_{ex}$} for this QD is calculated as 2.874, while the value is 2.946 from PL result.
This difference is attributed to the tunability of \emph{g$_{ex}$} by electric field \cite{sheng2010g}. Here, for the PL spectra, the electric field is 16 kV/cm at 0.4 V; while for the PC, it is about 38.9 kV/cm at zero magnetic field. This weak tunability of only about 2.5$\%$ is reasonable for self-assemble InAs QDs, comparing with flush-overgrown QDs having large tunability of exciton g-factor \cite{klotz2010observation}. The large \emph{g$_{ex}$} tuning has been observed in quantum dot molecules \cite{Doty2006} or quantum dots of high aspect ratio between the height and the base \cite{sheng2010g}. With such systems, the electron and hole wavefunction distribution could be modified easily with electric field \cite{klotz2010observation}, particularly for the hole wavefunction distribution \cite{Jovanov2011,sheng2010g}, which is not happening in the normal InAs QDs in our case. Noting that the PC signals have a wide electric field range at high magnetic field, so the extracted \emph{g$_{ex}$} is an approximate value from PC.

The energy shifts also bring changes on the PC signal's peak amplitude and linewidth, which is caused by the combined effect of both external electric and magnetic field.
With the increasing of electric field, the electron and the hole tunnel out of the QDs faster, which induces the linewidth broadening and the PC peak amplitude increase \cite{beham2001nonlinear,mar2011voltage,mar2011electrically}; meanwhile, the magnetic field's confinement for carriers in the QDs decrease the tunneling rates, resulting in narrower linewidth and smaller PC peak amplitude \cite{patan2002probing,godden2012fast}. These two opposite interactions both account for the linewidth and the PC amplitude changes in Fig.~\ref{fig:2}(b).

When the magnetic field is applied in the QD plane (Voigt geometry), the phenomenon is quite different. Instead of the circular polarization in Faraday geometry, the splitting peaks for different angular momenta are perpendicular linearly polarized in Voigt geometry. Fig.~\ref{fig:3}(a) shows the polarization-resolved PL spectra of X$^0$ and XX with the magnetic field range from 0-4 T in Voigt geometry at 0.4 V. Solid lines and dotted lines represent the horizontal and vertical linear polarization of the PL spectra, respectively. The diamagnetic effect and the Zeeman splitting is much weaker than those in Faraday geometry. The bright exciton splitting with in-plane magnetic field can be described as \emph{E}$\approx$ $\delta$$_1$+\emph{K}\emph{B}$^2$ \cite{stevenson2006magnetic}, where $\delta$$_1$ is the FSS for zero field, i.e. 18 $\mu$¦ÌeV for this dot. Here, the effective FSS increases with the in-plane magnetic field, as marked by dash-dot lines for the bright exciton emission of X$^0$ and XX in Fig.~\ref{fig:3}(a). At 4 T, the splitting between two bright exciton emissions is about 26 $\mu$eV.

\begin{figure}
\includegraphics[scale=0.8]{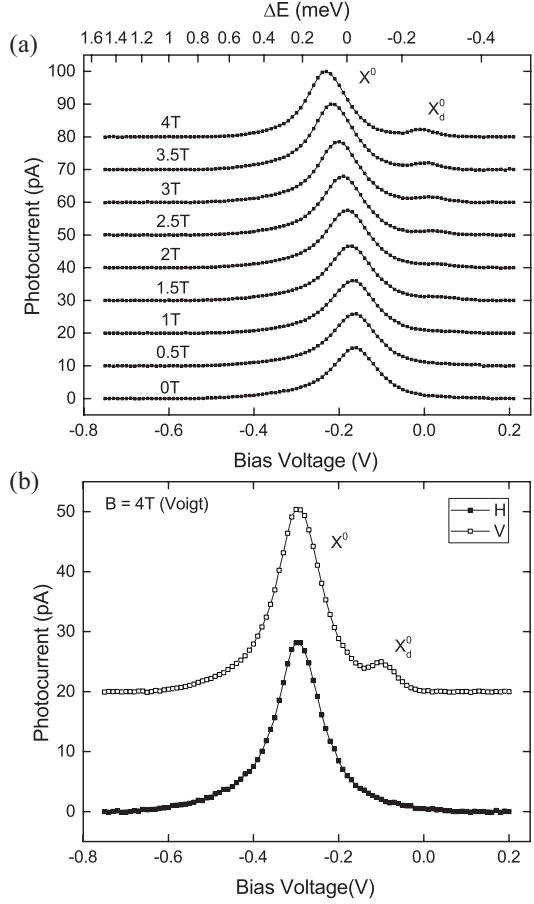}
\caption{\label{fig:4}(a) PC spectra of X$^0$ with magnetic field from 0 T to 4 T in Voigt geometry. The pumping laser energy is \emph{E}$_{Laser}$=1377.21 meV. The top x-axis represents the energy shift comparing with the X$^0$ transition energy at zero magnetic field. The dark exciton states X$^0_d$ can be observed from 1.5 T. (b) The polarization-resolved PC spectra of X$^0$ at 4 T in Voigt geometry. The dark exciton state only appears when pumping with a V-polarized laser. In order to observe more clearly, the pumping laser energy was tuned to \emph{E}$_{Laser}$=1377.03 meV, and the pumping power was also optimised. The spectra are shifted for clarity.
}
\end{figure}

The most striking feature of the excitons of QDs in Voigt geometry is the emergence of dark exciton states, which are optically forbidden normally with angular momentum of $\left|{M}\right|$ = 2. The in-plane magnetic field couples the bright and the dark states with destroying the rotational symmetry, resulting in the spectroscopic access of dark excitons. As shown in Fig.~\ref{fig:3}(a) labeled as X$^0_d$ and XX$_d$, the dark exciton peaks of X$^0$ and XX emerge when the magnetic field reaches 1.5 T, and become stronger with further increase of magnetic field. Using a diamagnetic field induced quadratic fitting, the dark state energy of X$^0$ and XX at zero magnetic field are extrapolated, as shown in Fig.~\ref{fig:3}(a). The bright and dark exciton energy difference is calculated as 0.258 meV, which is comparable with other works \cite{stevenson2006magnetic,schwartz2015deterministic}. But here only one branch of the two dark exciton states is observed similar to what have been reported previously \cite{stevenson2004time,stevenson2006magnetic}, which might due to the QDs asymmetry induced large anisotropy between the dipole moments of the two dark exciton states \cite{Zielinski2015}. Even the 4 T in-plane magnetic field can not destroy it.

The PC spectra of X$^0$ in Voigt geometry is collected by pumping with circular polarized light, as shown in Fig.~\ref{fig:4}. A dark exciton peak is observed at 1.5 T with the absorption energy lower than the bright exciton peak, consistent with the PL spectra. As the recombination of electron and hole in PL, here the optical transition of the dark exciton happens with absorbing a photon due to coupling of bright and dark exction states. The stronger coupling at high magnetic field in Voigt geometry results in an increase of PC amplitude.
The dark exciton has also been observed through PC spectrum by sweeping laser energy with coherent pulse pumping \cite{godden2012fast}. But here the resolution is much higher because of the narrow bandwidth ($\sim$1MHz) of CW laser comparing with the 0.2 meV bandwidth of the short pulse. Because of the circular polarization excitation, the PC peaks in Fig.~\ref{fig:4}(a) consist of both the polarization components. Then the polarization-resolved PC spectroscopy pumping by linear polarization laser was performed at 4 T in Voigt geometry. The result in Fig.~\ref{fig:4}(b) also shows the lack of H-polarized peak of dark exciton states corresponding well to the PL spectra, which further proves that this small peak is one of the dark exciton states.

\section{\label{sec:level1}Conclusion}

In conclusion, we have demonstrated the PL and PC spectroscopy of a single InAs/GaAs QD with applying magnetic field in Faraday and Voigt geometries. The high-resolution PC spectra of X$^0$ were obtained by sweeping the X$^0$ transition energy to match the fixed narrow-bandwidth laser via QCSE. Besides the typical diamagnetic effect and the Zeeman splitting, the dark exciton state was observed in Voigt geometry due to the mixing of bright and dark states. The PC spectra shows the similar behavior with the PL results. Because of the narrow bandwidth of the CW laser, the PC spectrum can have very narrow linewidth with a suitable device structure \cite{stufler2004power,stufler2005quantum} with high precision. This proves that the PC spectroscopy can be a powerful method to probe the dark exciton states in single QDs, as potential candidates to implement spin qubits with a long coherence time.

\begin{acknowledgments}
This work was supported by the National Basic Research Program of China under Grant No. 2014CB921003; the National Natural Science Foundation of China under Grant No.11721404, 91436101 and 61675228; the Strategic Priority Research Program of the Chinese Academy of Sciences under Grant No. XDB07030200 and XDPB0803, and the CAS Interdisciplinary Innovation Team.
\end{acknowledgments}

\appendix




\begin{thebibliography}{34}%
\makeatletter
\providecommand \@ifxundefined [1]{%
 \@ifx{#1\undefined}
}%
\providecommand \@ifnum [1]{%
 \ifnum #1\expandafter \@firstoftwo
 \else \expandafter \@secondoftwo
 \fi
}%
\providecommand \@ifx [1]{%
 \ifx #1\expandafter \@firstoftwo
 \else \expandafter \@secondoftwo
 \fi
}%
\providecommand \natexlab [1]{#1}%
\providecommand \enquote  [1]{#1}%
\providecommand \bibnamefont  [1]{#1}%
\providecommand \bibfnamefont [1]{#1}%
\providecommand \citenamefont [1]{#1}%
\providecommand \href@noop [0]{\@secondoftwo}%
\providecommand \href [0]{\begingroup \@sanitize@url \@href}%
\providecommand \@href[1]{\@@startlink{#1}\@@href}%
\providecommand \@@href[1]{\endgroup#1\@@endlink}%
\providecommand \@sanitize@url [0]{\catcode `\\12\catcode `\$12\catcode
  `\&12\catcode `\#12\catcode `\^12\catcode `\_12\catcode `\%12\relax}%
\providecommand \@@startlink[1]{}%
\providecommand \@@endlink[0]{}%
\providecommand \url  [0]{\begingroup\@sanitize@url \@url }%
\providecommand \@url [1]{\endgroup\@href {#1}{\urlprefix }}%
\providecommand \urlprefix  [0]{URL }%
\providecommand \Eprint [0]{\href }%
\providecommand \doibase [0]{http://dx.doi.org/}%
\providecommand \selectlanguage [0]{\@gobble}%
\providecommand \bibinfo  [0]{\@secondoftwo}%
\providecommand \bibfield  [0]{\@secondoftwo}%
\providecommand \translation [1]{[#1]}%
\providecommand \BibitemOpen [0]{}%
\providecommand \bibitemStop [0]{}%
\providecommand \bibitemNoStop [0]{.\EOS\space}%
\providecommand \EOS [0]{\spacefactor3000\relax}%
\providecommand \BibitemShut  [1]{\csname bibitem#1\endcsname}%
\let\auto@bib@innerbib\@empty
\bibitem [{\citenamefont {Loss}\ and\ \citenamefont
  {DiVincenzo}(1998)}]{loss1998quantum}%
  \BibitemOpen
  \bibfield  {author} {\bibinfo {author} {\bibfnamefont {D.}~\bibnamefont
  {Loss}}\ and\ \bibinfo {author} {\bibfnamefont {D.~P.}\ \bibnamefont
  {DiVincenzo}},\ }\bibfield  {title} {\enquote {\bibinfo {title} {Quantum
  computation with quantum dots},}\ }\href {\doibase 10.1103/PhysRevA.57.120} {\bibfield
  {journal} {\bibinfo  {journal} {Phys. Rev. A}\ }\textbf {\bibinfo {volume}
  {57}},\ \bibinfo {pages} {120} (\bibinfo {year} {1998})}\BibitemShut
  {NoStop}%
\bibitem [{\citenamefont {Imamoglu}\ \emph {et~al.}(1999)\citenamefont
  {Imamoglu}, \citenamefont {Awschalom}, \citenamefont {Burkard}, \citenamefont
  {DiVincenzo}, \citenamefont {Loss}, \citenamefont {Sherwin},\ and\
  \citenamefont {Small}}]{imamog1999quantum}%
  \BibitemOpen
  \bibfield  {author} {\bibinfo {author} {\bibfnamefont {A.}~\bibnamefont
  {Imamoglu}}, \bibinfo {author} {\bibfnamefont {D.~D.}\ \bibnamefont
  {Awschalom}}, \bibinfo {author} {\bibfnamefont {G.}~\bibnamefont {Burkard}},
  \bibinfo {author} {\bibfnamefont {D.~P.}\ \bibnamefont {DiVincenzo}},
  \bibinfo {author} {\bibfnamefont {D.}~\bibnamefont {Loss}}, \bibinfo {author}
  {\bibfnamefont {M.}~\bibnamefont {Sherwin}}, \ and\ \bibinfo {author}
  {\bibfnamefont {A.}~\bibnamefont {Small}},\ }\bibfield {title} {\enquote
  {\bibinfo {title} {Quantum information processing using quantum dot spins and
  cavity QED},}\ }\href {\doibase
  10.1103/PhysRevLett.83.4204} {\bibfield  {journal} {\bibinfo  {journal}
  {Phys. Rev. Lett.}\ }\textbf {\bibinfo {volume} {83}},\ \bibinfo {pages}
  {4204} (\bibinfo {year} {1999})}\BibitemShut {NoStop}%
\bibitem [{\citenamefont {Xu}\ \emph {et~al.}(2007)\citenamefont {Xu},
  \citenamefont {Wu}, \citenamefont {Sun}, \citenamefont {Huang}, \citenamefont
  {Cheng}, \citenamefont {Steel}, \citenamefont {Bracker}, \citenamefont
  {Gammon}, \citenamefont {Emary},\ and\ \citenamefont {Sham}}]{xu2007fast}%
  \BibitemOpen
  \bibfield  {author} {\bibinfo {author} {\bibfnamefont {X.}~\bibnamefont
  {Xu}}, \bibinfo {author} {\bibfnamefont {Y.}~\bibnamefont {Wu}}, \bibinfo
  {author} {\bibfnamefont {B.}~\bibnamefont {Sun}}, \bibinfo {author}
  {\bibfnamefont {Q.}~\bibnamefont {Huang}}, \bibinfo {author} {\bibfnamefont
  {J.}~\bibnamefont {Cheng}}, \bibinfo {author} {\bibfnamefont {D.~G.}\
  \bibnamefont {Steel}}, \bibinfo {author} {\bibfnamefont {A.~S.}\ \bibnamefont
  {Bracker}}, \bibinfo {author} {\bibfnamefont {D.}~\bibnamefont {Gammon}},
  \bibinfo {author} {\bibfnamefont {C.}~\bibnamefont {Emary}}, \ and\ \bibinfo
  {author} {\bibfnamefont {L.~J.}\ \bibnamefont {Sham}},\ }\bibfield  {title} {\enquote {\bibinfo {title} {Fast spin state
  initialization in a singly charged InAs-GaAs quantum dot by optical
  cooling},}\ }\href {\doibase
  10.1103/PhysRevLett.99.097401} {\bibfield  {journal} {\bibinfo  {journal}
  {Phys. Rev. Lett.}\ }\textbf {\bibinfo {volume} {99}},\ \bibinfo {pages}
  {097401} (\bibinfo {year} {2007})}\BibitemShut {NoStop}%
\bibitem [{\citenamefont {Berezovsky}\ \emph {et~al.}(2008)\citenamefont
  {Berezovsky}, \citenamefont {Mikkelsen}, \citenamefont {Stoltz},
  \citenamefont {Coldren},\ and\ \citenamefont {Awschalom}}]{Berezovsky349}%
  \BibitemOpen
  \bibfield  {author} {\bibinfo {author} {\bibfnamefont {J.}~\bibnamefont
  {Berezovsky}}, \bibinfo {author} {\bibfnamefont {M.~H.}\ \bibnamefont
  {Mikkelsen}}, \bibinfo {author} {\bibfnamefont {N.~G.}\ \bibnamefont
  {Stoltz}}, \bibinfo {author} {\bibfnamefont {L.~A.}\ \bibnamefont {Coldren}},
  \ and\ \bibinfo {author} {\bibfnamefont {D.~D.}\ \bibnamefont {Awschalom}},\
  }\bibfield  {title} {\enquote {\bibinfo {title} {Picosecond coherent optical
  manipulation of a single electron spin in a quantum dot},}\ }\href {\doibase 10.1126/science.1154798} {\bibfield  {journal} {\bibinfo
  {journal} {Science}\ }\textbf {\bibinfo {volume} {320}},\ \bibinfo {pages}
  {349} (\bibinfo {year} {2008})}\BibitemShut {NoStop}%
\bibitem [{\citenamefont {Press}\ \emph {et~al.}(2008)\citenamefont {Press},
  \citenamefont {Ladd}, \citenamefont {Zhang},\ and\ \citenamefont
  {Yamamoto}}]{press2008}%
  \BibitemOpen
  \bibfield  {author} {\bibinfo {author} {\bibfnamefont {D.}~\bibnamefont
  {Press}}, \bibinfo {author} {\bibfnamefont {T.~D.}\ \bibnamefont {Ladd}},
  \bibinfo {author} {\bibfnamefont {B.}~\bibnamefont {Zhang}}, \ and\ \bibinfo
  {author} {\bibfnamefont {Y.}~\bibnamefont {Yamamoto}},\ }\bibfield  {title} {\enquote {\bibinfo {title} {Complete
  quantum control of a single quantum dot spin using ultrafast optical
  pulses},}\ }\href
  {http://www.nature.com/nature/journal/v456/n7219/full/nature07530.html}
  {\bibfield  {journal} {\bibinfo  {journal} {Nature}\ }\textbf {\bibinfo
  {volume} {456}},\ \bibinfo {pages} {218} (\bibinfo {year}
  {2008})}\BibitemShut {NoStop}%
\bibitem [{\citenamefont {Brunner}\ \emph {et~al.}(2009)\citenamefont
  {Brunner}, \citenamefont {Gerardot}, \citenamefont {Dalgarno}, \citenamefont
  {W{\"u}st}, \citenamefont {Karrai}, \citenamefont {Stoltz}, \citenamefont
  {Petroff},\ and\ \citenamefont {Warburton}}]{brunner2009coherent}%
  \BibitemOpen
  \bibfield  {author} {\bibinfo {author} {\bibfnamefont {D.}~\bibnamefont
  {Brunner}}, \bibinfo {author} {\bibfnamefont {B.~D.}\ \bibnamefont
  {Gerardot}}, \bibinfo {author} {\bibfnamefont {P.~A.}\ \bibnamefont
  {Dalgarno}}, \bibinfo {author} {\bibfnamefont {G.}~\bibnamefont {W{\"u}st}},
  \bibinfo {author} {\bibfnamefont {K.}~\bibnamefont {Karrai}}, \bibinfo
  {author} {\bibfnamefont {N.~G.}\ \bibnamefont {Stoltz}}, \bibinfo {author}
  {\bibfnamefont {P.~M.}\ \bibnamefont {Petroff}}, \ and\ \bibinfo {author}
  {\bibfnamefont {R.~J.}\ \bibnamefont {Warburton}},\ }\bibfield  {title} {\enquote {\bibinfo {title}
  {A coherent single-hole spin in a semiconductor},}\ }\href {\doibase
  10.1126/science.1173684} {\bibfield  {journal} {\bibinfo  {journal}
  {Science}\ }\textbf {\bibinfo {volume} {325}},\ \bibinfo {pages} {70}
  (\bibinfo {year} {2009})}\BibitemShut {NoStop}%
\bibitem [{\citenamefont {Kim}\ \emph {et~al.}(2010)\citenamefont {Kim},
  \citenamefont {Truex}, \citenamefont {Xu}, \citenamefont {Sun}, \citenamefont
  {Steel}, \citenamefont {Bracker}, \citenamefont {Gammon},\ and\ \citenamefont
  {Sham}}]{Kim2010Fast}%
  \BibitemOpen
  \bibfield  {author} {\bibinfo {author} {\bibfnamefont {E.~D.}\ \bibnamefont
  {Kim}}, \bibinfo {author} {\bibfnamefont {K.}~\bibnamefont {Truex}}, \bibinfo
  {author} {\bibfnamefont {X.}~\bibnamefont {Xu}}, \bibinfo {author}
  {\bibfnamefont {B.}~\bibnamefont {Sun}}, \bibinfo {author} {\bibfnamefont
  {D.~G.}\ \bibnamefont {Steel}}, \bibinfo {author} {\bibfnamefont {A.~S.}\
  \bibnamefont {Bracker}}, \bibinfo {author} {\bibfnamefont {D.}~\bibnamefont
  {Gammon}}, \ and\ \bibinfo {author} {\bibfnamefont {L.~J.}\ \bibnamefont
  {Sham}},\ }\bibfield  {title} {\enquote
  {\bibinfo {title} {Fast spin rotations by optically controlled geometric
  phases in a charge-tunable InAs quantum dot},}\ }\href {\doibase 10.1103/PhysRevLett.104.167401} {\bibfield
  {journal} {\bibinfo  {journal} {Phys. Rev. Lett.}\ }\textbf {\bibinfo
  {volume} {104}},\ \bibinfo {pages} {167401} (\bibinfo {year}
  {2010})}\BibitemShut {NoStop}%
\bibitem [{\citenamefont {Godden}\ \emph
  {et~al.}(2012{\natexlab{a}})\citenamefont {Godden}, \citenamefont {Quilter},
  \citenamefont {Ramsay}, \citenamefont {Wu}, \citenamefont {Brereton},
  \citenamefont {Boyle}, \citenamefont {Luxmoore}, \citenamefont
  {Puebla-Nunez}, \citenamefont {Fox},\ and\ \citenamefont
  {Skolnick}}]{godden2012coherent}%
  \BibitemOpen
  \bibfield  {author} {\bibinfo {author} {\bibfnamefont {T.~M.}\ \bibnamefont
  {Godden}}, \bibinfo {author} {\bibfnamefont {J.~H.}\ \bibnamefont {Quilter}},
  \bibinfo {author} {\bibfnamefont {A.~J.}\ \bibnamefont {Ramsay}}, \bibinfo
  {author} {\bibfnamefont {Y.}~\bibnamefont {Wu}}, \bibinfo {author}
  {\bibfnamefont {P.}~\bibnamefont {Brereton}}, \bibinfo {author}
  {\bibfnamefont {S.~J.}\ \bibnamefont {Boyle}}, \bibinfo {author}
  {\bibfnamefont {I.~J.}\ \bibnamefont {Luxmoore}}, \bibinfo {author}
  {\bibfnamefont {J.}~\bibnamefont {Puebla-Nunez}}, \bibinfo {author}
  {\bibfnamefont {A.~M.}\ \bibnamefont {Fox}}, \ and\ \bibinfo {author}
  {\bibfnamefont {M.~S.}\ \bibnamefont {Skolnick}},\ }\bibfield  {title}
  {\enquote {\bibinfo {title} {Coherent optical control of the spin of a single
  hole in an $\mathrm{InAs}/\mathrm{GaAs}$ quantum dot},}\ }\href {\doibase
  10.1103/PhysRevLett.108.017402} {\bibfield  {journal} {\bibinfo  {journal}
  {Phys. Rev. Lett.}\ }\textbf {\bibinfo {volume} {108}},\ \bibinfo {pages}
  {017402} (\bibinfo {year} {2012}{\natexlab{a}})}\BibitemShut {NoStop}%
\bibitem [{\citenamefont {Warburton}(2013)}]{warburton2013single}%
  \BibitemOpen
  \bibfield  {author} {\bibinfo {author} {\bibfnamefont {R.~J.}\ \bibnamefont
  {Warburton}},\ }\bibfield  {title} {\enquote {\bibinfo {title}
  {Single spins in self-assembled quantum dots},}\ }\href
  {http://www.nature.com/nmat/journal/v12/n6/full/nmat3585.html?foxtrotcallback=true}
  {\bibfield  {journal} {\bibinfo  {journal} {Nat. Mater}\ }\textbf {\bibinfo
  {volume} {12}},\ \bibinfo {pages} {483} (\bibinfo {year} {2013})}\BibitemShut
  {NoStop}%
\bibitem [{\citenamefont {Bayer}\ \emph {et~al.}(2002)\citenamefont {Bayer},
  \citenamefont {Ortner}, \citenamefont {Stern}, \citenamefont {Kuther},
  \citenamefont {Gorbunov}, \citenamefont {Forchel}, \citenamefont {Hawrylak},
  \citenamefont {Fafard}, \citenamefont {Hinzer}, \citenamefont {Reinecke},
  \citenamefont {Walck}, \citenamefont {Reithmaier}, \citenamefont {Klopf},\
  and\ \citenamefont {Sch\"afer}}]{bayer2002fine}%
  \BibitemOpen
  \bibfield  {author} {\bibinfo {author} {\bibfnamefont {M.}~\bibnamefont
  {Bayer}}, \bibinfo {author} {\bibfnamefont {G.}~\bibnamefont {Ortner}},
  \bibinfo {author} {\bibfnamefont {O.}~\bibnamefont {Stern}}, \bibinfo
  {author} {\bibfnamefont {A.}~\bibnamefont {Kuther}}, \bibinfo {author}
  {\bibfnamefont {A.~A.}\ \bibnamefont {Gorbunov}}, \bibinfo {author}
  {\bibfnamefont {A.}~\bibnamefont {Forchel}}, \bibinfo {author} {\bibfnamefont
  {P.}~\bibnamefont {Hawrylak}}, \bibinfo {author} {\bibfnamefont
  {S.}~\bibnamefont {Fafard}}, \bibinfo {author} {\bibfnamefont
  {K.}~\bibnamefont {Hinzer}}, \bibinfo {author} {\bibfnamefont {T.~L.}\
  \bibnamefont {Reinecke}}, \bibinfo {author} {\bibfnamefont {S.~N.}\
  \bibnamefont {Walck}}, \bibinfo {author} {\bibfnamefont {J.~P.}\ \bibnamefont
  {Reithmaier}}, \bibinfo {author} {\bibfnamefont {F.}~\bibnamefont {Klopf}}, \
  and\ \bibinfo {author} {\bibfnamefont {F.}~\bibnamefont {Sch\"afer}},\ }\bibfield  {title} {\enquote {\bibinfo {title} {Fine structure of neutral
  and charged excitons in self-assembled In(Ga)As/(Al)GaAs quantum dots},}\
  }\href
  {\doibase 10.1103/PhysRevB.65.195315} {\bibfield  {journal} {\bibinfo
  {journal} {Phys. Rev. B}\ }\textbf {\bibinfo {volume} {65}},\ \bibinfo
  {pages} {195315} (\bibinfo {year} {2002})}\BibitemShut {NoStop}%
\bibitem [{\citenamefont {Poem}\ \emph {et~al.}(2010)\citenamefont {Poem},
  \citenamefont {Kodriano}, \citenamefont {Tradonsky}, \citenamefont {Lindner},
  \citenamefont {Gerardot}, \citenamefont {Petroff},\ and\ \citenamefont
  {Gershoni}}]{poem2010accessing}%
  \BibitemOpen
  \bibfield  {author} {\bibinfo {author} {\bibfnamefont {E.}~\bibnamefont
  {Poem}}, \bibinfo {author} {\bibfnamefont {Y.}~\bibnamefont {Kodriano}},
  \bibinfo {author} {\bibfnamefont {C.}~\bibnamefont {Tradonsky}}, \bibinfo
  {author} {\bibfnamefont {N.}~\bibnamefont {Lindner}}, \bibinfo {author}
  {\bibfnamefont {B.}~\bibnamefont {Gerardot}}, \bibinfo {author}
  {\bibfnamefont {P.}~\bibnamefont {Petroff}}, \ and\ \bibinfo {author}
  {\bibfnamefont {D.}~\bibnamefont {Gershoni}},\ }\bibfield  {title} {\enquote
  {\bibinfo {title} {Accessing the dark exciton with light},}\ }\href
  {https://www.nature.com/nphys/journal/v6/n12/full/nphys1812.html} {\bibfield
  {journal} {\bibinfo  {journal} {Nat. Phys}\ }\textbf {\bibinfo {volume}
  {6}},\ \bibinfo {pages} {993} (\bibinfo {year} {2010})}\BibitemShut {NoStop}%
\bibitem [{\citenamefont {Schwartz}\ \emph {et~al.}(2015)\citenamefont
  {Schwartz}, \citenamefont {Schmidgall}, \citenamefont {Gantz}, \citenamefont
  {Cogan}, \citenamefont {Bordo}, \citenamefont {Don}, \citenamefont
  {Zielinski},\ and\ \citenamefont {Gershoni}}]{schwartz2015deterministic}%
  \BibitemOpen
  \bibfield  {author} {\bibinfo {author} {\bibfnamefont {I.}~\bibnamefont
  {Schwartz}}, \bibinfo {author} {\bibfnamefont {E.~R.}\ \bibnamefont
  {Schmidgall}}, \bibinfo {author} {\bibfnamefont {L.}~\bibnamefont {Gantz}},
  \bibinfo {author} {\bibfnamefont {D.}~\bibnamefont {Cogan}}, \bibinfo
  {author} {\bibfnamefont {E.}~\bibnamefont {Bordo}}, \bibinfo {author}
  {\bibfnamefont {Y.}~\bibnamefont {Don}}, \bibinfo {author} {\bibfnamefont
  {M.}~\bibnamefont {Zielinski}}, \ and\ \bibinfo {author} {\bibfnamefont
  {D.}~\bibnamefont {Gershoni}},\ }\bibfield  {title} {\enquote {\bibinfo
  {title} {Deterministic writing and control of the dark exciton spin using
  single short optical pulses},}\ }\href {\doibase 10.1103/PhysRevX.5.011009}
  {\bibfield  {journal} {\bibinfo  {journal} {Phys. Rev. X}\ }\textbf {\bibinfo
  {volume} {5}},\ \bibinfo {pages} {011009} (\bibinfo {year}
  {2015})}\BibitemShut {NoStop}%
\bibitem [{\citenamefont {Tsai}\ \emph {et~al.}(2008)\citenamefont {Tsai},
  \citenamefont {Lin}, \citenamefont {Lin}, \citenamefont {Lin}, \citenamefont
  {Wang}, \citenamefont {Lo}, \citenamefont {Cheng}, \citenamefont {Lee},\ and\
  \citenamefont {Chang}}]{Tsai2008Diamagnetic}%
  \BibitemOpen
  \bibfield  {author} {\bibinfo {author} {\bibfnamefont {M.-F.}\ \bibnamefont
  {Tsai}}, \bibinfo {author} {\bibfnamefont {H.}~\bibnamefont {Lin}}, \bibinfo
  {author} {\bibfnamefont {C.-H.}\ \bibnamefont {Lin}}, \bibinfo {author}
  {\bibfnamefont {S.-D.}\ \bibnamefont {Lin}}, \bibinfo {author} {\bibfnamefont
  {S.-Y.}\ \bibnamefont {Wang}}, \bibinfo {author} {\bibfnamefont {M.-C.}\
  \bibnamefont {Lo}}, \bibinfo {author} {\bibfnamefont {S.-J.}\ \bibnamefont
  {Cheng}}, \bibinfo {author} {\bibfnamefont {M.-C.}\ \bibnamefont {Lee}}, \
  and\ \bibinfo {author} {\bibfnamefont {W.-H.}\ \bibnamefont {Chang}},\ }\bibfield  {title}
  {\enquote {\bibinfo {title} {Diamagnetic response of exciton complexes in
  semiconductor quantum dots},}\ }\href
  {\doibase 10.1103/PhysRevLett.101.267402} {\bibfield  {journal} {\bibinfo
  {journal} {Phys. Rev. Lett.}\ }\textbf {\bibinfo {volume} {101}},\ \bibinfo
  {pages} {267402} (\bibinfo {year} {2008})}\BibitemShut {NoStop}%
\bibitem [{\citenamefont {Cao}\ \emph {et~al.}(2016)\citenamefont {Cao},
  \citenamefont {Tang}, \citenamefont {Sun}, \citenamefont {Peng},
  \citenamefont {Gao}, \citenamefont {Zhao}, \citenamefont {Qian},
  \citenamefont {Sun}, \citenamefont {Ali}, \citenamefont {Shao}, \citenamefont
  {Wu}, \citenamefont {Song}, \citenamefont {Williams}, \citenamefont {Sheng},
  \citenamefont {Jin},\ and\ \citenamefont {Xu}}]{Cao2016}%
  \BibitemOpen
  \bibfield  {author} {\bibinfo {author} {\bibfnamefont {S.}~\bibnamefont
  {Cao}}, \bibinfo {author} {\bibfnamefont {J.}~\bibnamefont {Tang}}, \bibinfo
  {author} {\bibfnamefont {Y.}~\bibnamefont {Sun}}, \bibinfo {author}
  {\bibfnamefont {K.}~\bibnamefont {Peng}}, \bibinfo {author} {\bibfnamefont
  {Y.}~\bibnamefont {Gao}}, \bibinfo {author} {\bibfnamefont {Y.}~\bibnamefont
  {Zhao}}, \bibinfo {author} {\bibfnamefont {C.}~\bibnamefont {Qian}}, \bibinfo
  {author} {\bibfnamefont {S.}~\bibnamefont {Sun}}, \bibinfo {author}
  {\bibfnamefont {H.}~\bibnamefont {Ali}}, \bibinfo {author} {\bibfnamefont
  {Y.}~\bibnamefont {Shao}}, \bibinfo {author} {\bibfnamefont {S.}~\bibnamefont
  {Wu}}, \bibinfo {author} {\bibfnamefont {F.}~\bibnamefont {Song}}, \bibinfo
  {author} {\bibfnamefont {D.~A.}\ \bibnamefont {Williams}}, \bibinfo {author}
  {\bibfnamefont {W.}~\bibnamefont {Sheng}}, \bibinfo {author} {\bibfnamefont
  {K.}~\bibnamefont {Jin}}, \ and\ \bibinfo {author} {\bibfnamefont
  {X.}~\bibnamefont {Xu}},\ }\bibfield  {title} {\enquote
  {\bibinfo {title} {Observation of coupling between zero- and two-dimensional
  semiconductor systems based on anomalous diamagnetic effects},}\ }\href {\doibase 10.1007/s12274-015-0910-z}
  {\bibfield  {journal} {\bibinfo  {journal} {Nano Research}\ }\textbf
  {\bibinfo {volume} {9}},\ \bibinfo {pages} {306} (\bibinfo {year}
  {2016})}\BibitemShut {NoStop}%
\bibitem [{\citenamefont {Zrenner}\ \emph {et~al.}(2002)\citenamefont
  {Zrenner}, \citenamefont {Beham}, \citenamefont {Stufler}, \citenamefont
  {Findeis}, \citenamefont {Bichler},\ and\ \citenamefont
  {Abstreiter}}]{zrenner2002coherent}%
  \BibitemOpen
  \bibfield  {author} {\bibinfo {author} {\bibfnamefont {A.}~\bibnamefont
  {Zrenner}}, \bibinfo {author} {\bibfnamefont {E.}~\bibnamefont {Beham}},
  \bibinfo {author} {\bibfnamefont {S.}~\bibnamefont {Stufler}}, \bibinfo
  {author} {\bibfnamefont {F.}~\bibnamefont {Findeis}}, \bibinfo {author}
  {\bibfnamefont {M.}~\bibnamefont {Bichler}}, \ and\ \bibinfo {author}
  {\bibfnamefont {G.}~\bibnamefont {Abstreiter}},\ }\bibfield  {title} {\enquote
  {\bibinfo {title} {Coherent properties of a two-level system based on a
  quantum-dot photodiode},}\ }\href
  {http://dx.doi.org/10.1038/nature00912} {\bibfield  {journal} {\bibinfo
  {journal} {Nature}\ }\textbf {\bibinfo {volume} {418}},\ \bibinfo {pages}
  {612} (\bibinfo {year} {2002})}\BibitemShut {NoStop}%
\bibitem [{\citenamefont {Stufler}\ \emph {et~al.}(2006)\citenamefont
  {Stufler}, \citenamefont {Machnikowski}, \citenamefont {Ester}, \citenamefont
  {Bichler}, \citenamefont {Axt}, \citenamefont {Kuhn},\ and\ \citenamefont
  {Zrenner}}]{stufler2006two}%
  \BibitemOpen
  \bibfield  {author} {\bibinfo {author} {\bibfnamefont {S.}~\bibnamefont
  {Stufler}}, \bibinfo {author} {\bibfnamefont {P.}~\bibnamefont
  {Machnikowski}}, \bibinfo {author} {\bibfnamefont {P.}~\bibnamefont {Ester}},
  \bibinfo {author} {\bibfnamefont {M.}~\bibnamefont {Bichler}}, \bibinfo
  {author} {\bibfnamefont {V.~M.}\ \bibnamefont {Axt}}, \bibinfo {author}
  {\bibfnamefont {T.}~\bibnamefont {Kuhn}}, \ and\ \bibinfo {author}
  {\bibfnamefont {A.}~\bibnamefont {Zrenner}},\ }\bibfield  {title} {\enquote
  {\bibinfo {title} {Two-photon rabi oscillations in a single
  ${\mathrm{In}}_{x}{\mathrm{Ga}}_{1\ensuremath{-}x}\mathrm{As}/\mathrm{Ga}\mathrm{As}$
  quantum dot},}\ }\href {\doibase
  10.1103/PhysRevB.73.125304} {\bibfield  {journal} {\bibinfo  {journal} {Phys.
  Rev. B}\ }\textbf {\bibinfo {volume} {73}},\ \bibinfo {pages} {125304}
  (\bibinfo {year} {2006})}\BibitemShut {NoStop}%
\bibitem [{\citenamefont {Ramsay}\ \emph {et~al.}(2008)\citenamefont {Ramsay},
  \citenamefont {Boyle}, \citenamefont {Kolodka}, \citenamefont {Oliveira},
  \citenamefont {Skiba-Szymanska}, \citenamefont {Liu}, \citenamefont
  {Hopkinson}, \citenamefont {Fox},\ and\ \citenamefont
  {Skolnick}}]{Ramsay2008Fast}%
  \BibitemOpen
  \bibfield  {author} {\bibinfo {author} {\bibfnamefont {A.~J.}\ \bibnamefont
  {Ramsay}}, \bibinfo {author} {\bibfnamefont {S.~J.}\ \bibnamefont {Boyle}},
  \bibinfo {author} {\bibfnamefont {R.~S.}\ \bibnamefont {Kolodka}}, \bibinfo
  {author} {\bibfnamefont {J.~B.~B.}\ \bibnamefont {Oliveira}}, \bibinfo
  {author} {\bibfnamefont {J.}~\bibnamefont {Skiba-Szymanska}}, \bibinfo
  {author} {\bibfnamefont {H.~Y.}\ \bibnamefont {Liu}}, \bibinfo {author}
  {\bibfnamefont {M.}~\bibnamefont {Hopkinson}}, \bibinfo {author}
  {\bibfnamefont {A.~M.}\ \bibnamefont {Fox}}, \ and\ \bibinfo {author}
  {\bibfnamefont {M.~S.}\ \bibnamefont {Skolnick}},\ }\bibfield  {title}
  {\enquote {\bibinfo {title} {Fast optical preparation, control, and readout
  of a single quantum dot spin},}\ }\href {\doibase
  10.1103/PhysRevLett.100.197401} {\bibfield  {journal} {\bibinfo  {journal}
  {Phys. Rev. Lett.}\ }\textbf {\bibinfo {volume} {100}},\ \bibinfo {pages}
  {197401} (\bibinfo {year} {2008})}\BibitemShut {NoStop}%
\bibitem [{\citenamefont {Takagi}\ \emph {et~al.}(2008)\citenamefont {Takagi},
  \citenamefont {Nakaoka}, \citenamefont {Watanabe}, \citenamefont {Kumagai},\
  and\ \citenamefont {Arakawa}}]{takagi2008coherently}%
  \BibitemOpen
  \bibfield  {author} {\bibinfo {author} {\bibfnamefont {H.}~\bibnamefont
  {Takagi}}, \bibinfo {author} {\bibfnamefont {T.}~\bibnamefont {Nakaoka}},
  \bibinfo {author} {\bibfnamefont {K.}~\bibnamefont {Watanabe}}, \bibinfo
  {author} {\bibfnamefont {N.}~\bibnamefont {Kumagai}}, \ and\ \bibinfo
  {author} {\bibfnamefont {Y.}~\bibnamefont {Arakawa}},\ }\bibfield  {title} {\enquote {\bibinfo {title}
  {Coherently driven semiconductor quantum dot at a telecommunication
  wavelength},}\ }\href {\doibase
  10.1364/OE.16.013949} {\bibfield  {journal} {\bibinfo  {journal} {Opt.
  Express}\ }\textbf {\bibinfo {volume} {16}},\ \bibinfo {pages} {13949}
  (\bibinfo {year} {2008})}\BibitemShut {NoStop}%
\bibitem [{\citenamefont {Mar}\ \emph {et~al.}(2013)\citenamefont {Mar},
  \citenamefont {Baumberg}, \citenamefont {Xu}, \citenamefont {Irvine},
  \citenamefont {Stanley},\ and\ \citenamefont {Williams}}]{mar2013high}%
  \BibitemOpen
  \bibfield  {author} {\bibinfo {author} {\bibfnamefont {J.~D.}\ \bibnamefont
  {Mar}}, \bibinfo {author} {\bibfnamefont {J.~J.}\ \bibnamefont {Baumberg}},
  \bibinfo {author} {\bibfnamefont {X.~L.}\ \bibnamefont {Xu}}, \bibinfo
  {author} {\bibfnamefont {A.~C.}\ \bibnamefont {Irvine}}, \bibinfo {author}
  {\bibfnamefont {C.~R.}\ \bibnamefont {Stanley}}, \ and\ \bibinfo {author}
  {\bibfnamefont {D.~A.}\ \bibnamefont {Williams}},\ }\bibfield  {title}
  {\enquote {\bibinfo {title} {High-resolution photocurrent spectroscopy of the
  positive trion state in a single quantum dot},}\ }\href {\doibase
  10.1103/PhysRevB.87.155315} {\bibfield  {journal} {\bibinfo  {journal} {Phys.
  Rev. B}\ }\textbf {\bibinfo {volume} {87}},\ \bibinfo {pages} {155315}
  (\bibinfo {year} {2013})}\BibitemShut {NoStop}%
\bibitem [{\citenamefont {Mar}\ \emph {et~al.}(2017)\citenamefont {Mar},
  \citenamefont {Baumberg}, \citenamefont {Xu}, \citenamefont {Irvine},\ and\
  \citenamefont {Williams}}]{mar2017precise}%
  \BibitemOpen
  \bibfield  {author} {\bibinfo {author} {\bibfnamefont {J.~D.}\ \bibnamefont
  {Mar}}, \bibinfo {author} {\bibfnamefont {J.~J.}\ \bibnamefont {Baumberg}},
  \bibinfo {author} {\bibfnamefont {X.~L.}\ \bibnamefont {Xu}}, \bibinfo
  {author} {\bibfnamefont {A.~C.}\ \bibnamefont {Irvine}}, \ and\ \bibinfo
  {author} {\bibfnamefont {D.~A.}\ \bibnamefont {Williams}},\ }\bibfield
  {title} {\enquote {\bibinfo {title} {Precise measurements of the dipole
  moment and polarizability of the neutral exciton and positive trion in a
  single quantum dot},}\ }\href {\doibase
  10.1103/PhysRevB.95.201304} {\bibfield  {journal} {\bibinfo  {journal} {Phys.
  Rev. B}\ }\textbf {\bibinfo {volume} {95}},\ \bibinfo {pages} {201304}
  (\bibinfo {year} {2017})}\BibitemShut {NoStop}%
\bibitem [{\citenamefont {Mar}\ \emph {et~al.}(2014)\citenamefont {Mar},
  \citenamefont {Baumberg}, \citenamefont {Xu}, \citenamefont {Irvine},\ and\
  \citenamefont {Williams}}]{mar2014ultrafast}%
  \BibitemOpen
  \bibfield  {author} {\bibinfo {author} {\bibfnamefont {J.~D.}\ \bibnamefont
  {Mar}}, \bibinfo {author} {\bibfnamefont {J.~J.}\ \bibnamefont {Baumberg}},
  \bibinfo {author} {\bibfnamefont {X.}~\bibnamefont {Xu}}, \bibinfo {author}
  {\bibfnamefont {A.~C.}\ \bibnamefont {Irvine}}, \ and\ \bibinfo {author}
  {\bibfnamefont {D.~A.}\ \bibnamefont {Williams}},\ }\bibfield  {title} {\enquote {\bibinfo {title}
  {Ultrafast high-fidelity initialization of a quantum-dot spin qubit without
  magnetic fields},}\ }\href {\doibase
  10.1103/PhysRevB.90.241303} {\bibfield  {journal} {\bibinfo  {journal} {Phys.
  Rev. B}\ }\textbf {\bibinfo {volume} {90}},\ \bibinfo {pages} {241303}
  (\bibinfo {year} {2014})}\BibitemShut {NoStop}%
\bibitem [{\citenamefont {Tang}\ \emph {et~al.}(2014)\citenamefont {Tang},
  \citenamefont {Cao}, \citenamefont {Gao}, \citenamefont {Sun}, \citenamefont
  {Geng}, \citenamefont {Williams}, \citenamefont {Jin},\ and\ \citenamefont
  {Xu}}]{tang2014charge}%
  \BibitemOpen
  \bibfield  {author} {\bibinfo {author} {\bibfnamefont {J.}~\bibnamefont
  {Tang}}, \bibinfo {author} {\bibfnamefont {S.}~\bibnamefont {Cao}}, \bibinfo
  {author} {\bibfnamefont {Y.}~\bibnamefont {Gao}}, \bibinfo {author}
  {\bibfnamefont {Y.}~\bibnamefont {Sun}}, \bibinfo {author} {\bibfnamefont
  {W.}~\bibnamefont {Geng}}, \bibinfo {author} {\bibfnamefont {D.~A.}\
  \bibnamefont {Williams}}, \bibinfo {author} {\bibfnamefont {K.}~\bibnamefont
  {Jin}}, \ and\ \bibinfo {author} {\bibfnamefont {X.}~\bibnamefont {Xu}},\
  }\bibfield  {title} {\enquote
  {\bibinfo {title} {Charge state control in single InAs/GaAs quantum dots by
  external electric and magnetic fields},}\ }\href {\doibase 10.1063/1.4891828} {\bibfield  {journal} {\bibinfo
  {journal} {Appl. Phys. Lett}\ }\textbf {\bibinfo {volume} {105}},\ \bibinfo
  {pages} {041109} (\bibinfo {year} {2014})}\BibitemShut {NoStop}%
\bibitem [{\citenamefont {Sheng}(2010)}]{sheng2010g}%
  \BibitemOpen
  \bibfield  {author} {\bibinfo {author} {\bibfnamefont {W.}~\bibnamefont
  {Sheng}},\ }\bibfield  {title} {\enquote {\bibinfo {title} {g-factor tuning
  in self-assembled quantum dots},}\ }\href {\doibase 10.1063/1.3367707} {\bibfield  {journal}
  {\bibinfo  {journal} {Appl. Phys. Lett}\ }\textbf {\bibinfo {volume} {96}},\
  \bibinfo {pages} {133102} (\bibinfo {year} {2010})}\BibitemShut {NoStop}%
\bibitem [{\citenamefont {Klotz}\ \emph {et~al.}(2010)\citenamefont {Klotz},
  \citenamefont {Jovanov}, \citenamefont {Kierig}, \citenamefont {Clark},
  \citenamefont {Rudolph}, \citenamefont {Heiss}, \citenamefont {Bichler},
  \citenamefont {Abstreiter}, \citenamefont {Brandt},\ and\ \citenamefont
  {Finley}}]{klotz2010observation}%
  \BibitemOpen
  \bibfield  {author} {\bibinfo {author} {\bibfnamefont {F.}~\bibnamefont
  {Klotz}}, \bibinfo {author} {\bibfnamefont {V.}~\bibnamefont {Jovanov}},
  \bibinfo {author} {\bibfnamefont {J.}~\bibnamefont {Kierig}}, \bibinfo
  {author} {\bibfnamefont {E.~C.}\ \bibnamefont {Clark}}, \bibinfo {author}
  {\bibfnamefont {D.}~\bibnamefont {Rudolph}}, \bibinfo {author} {\bibfnamefont
  {D.}~\bibnamefont {Heiss}}, \bibinfo {author} {\bibfnamefont
  {M.}~\bibnamefont {Bichler}}, \bibinfo {author} {\bibfnamefont
  {G.}~\bibnamefont {Abstreiter}}, \bibinfo {author} {\bibfnamefont {M.~S.}\
  \bibnamefont {Brandt}}, \ and\ \bibinfo {author} {\bibfnamefont {J.~J.}\
  \bibnamefont {Finley}},\ }\bibfield  {title} {\enquote {\bibinfo {title}
  {Observation of an electrically tunable exciton g factor in InGaAs/GaAs
  quantum dots},}\ }\href {\doibase 10.1063/1.3309684} {\bibfield
  {journal} {\bibinfo  {journal} {Appl. Phys. Lett}\ }\textbf {\bibinfo
  {volume} {96}},\ \bibinfo {pages} {053113} (\bibinfo {year}
  {2010})}\BibitemShut {NoStop}%
\bibitem [{\citenamefont {Doty}\ \emph {et~al.}(2006)\citenamefont {Doty},
  \citenamefont {Scheibner}, \citenamefont {Ponomarev}, \citenamefont
  {Stinaff}, \citenamefont {Bracker}, \citenamefont {Korenev}, \citenamefont
  {Reinecke},\ and\ \citenamefont {Gammon}}]{Doty2006}%
  \BibitemOpen
  \bibfield  {author} {\bibinfo {author} {\bibfnamefont {M.~F.}\ \bibnamefont
  {Doty}}, \bibinfo {author} {\bibfnamefont {M.}~\bibnamefont {Scheibner}},
  \bibinfo {author} {\bibfnamefont {I.~V.}\ \bibnamefont {Ponomarev}}, \bibinfo
  {author} {\bibfnamefont {E.~A.}\ \bibnamefont {Stinaff}}, \bibinfo {author}
  {\bibfnamefont {A.~S.}\ \bibnamefont {Bracker}}, \bibinfo {author}
  {\bibfnamefont {V.~L.}\ \bibnamefont {Korenev}}, \bibinfo {author}
  {\bibfnamefont {T.~L.}\ \bibnamefont {Reinecke}}, \ and\ \bibinfo {author}
  {\bibfnamefont {D.}~\bibnamefont {Gammon}},\ }\bibfield  {title} {\enquote {\bibinfo {title}
  {Electrically tunable $g$ factors in quantum dot molecular spin states},}\ }\href {\doibase
  10.1103/PhysRevLett.97.197202} {\bibfield  {journal} {\bibinfo  {journal}
  {Phys. Rev. Lett.}\ }\textbf {\bibinfo {volume} {97}},\ \bibinfo {pages}
  {197202} (\bibinfo {year} {2006})}\BibitemShut {NoStop}%
\bibitem [{\citenamefont {Jovanov}\ \emph {et~al.}(2011)\citenamefont
  {Jovanov}, \citenamefont {Eissfeller}, \citenamefont {Kapfinger},
  \citenamefont {Clark}, \citenamefont {Klotz}, \citenamefont {Bichler},
  \citenamefont {Keizer}, \citenamefont {Koenraad}, \citenamefont
  {Abstreiter},\ and\ \citenamefont {Finley}}]{Jovanov2011}%
  \BibitemOpen
  \bibfield  {author} {\bibinfo {author} {\bibfnamefont {V.}~\bibnamefont
  {Jovanov}}, \bibinfo {author} {\bibfnamefont {T.}~\bibnamefont {Eissfeller}},
  \bibinfo {author} {\bibfnamefont {S.}~\bibnamefont {Kapfinger}}, \bibinfo
  {author} {\bibfnamefont {E.~C.}\ \bibnamefont {Clark}}, \bibinfo {author}
  {\bibfnamefont {F.}~\bibnamefont {Klotz}}, \bibinfo {author} {\bibfnamefont
  {M.}~\bibnamefont {Bichler}}, \bibinfo {author} {\bibfnamefont {J.~G.}\
  \bibnamefont {Keizer}}, \bibinfo {author} {\bibfnamefont {P.~M.}\
  \bibnamefont {Koenraad}}, \bibinfo {author} {\bibfnamefont {G.}~\bibnamefont
  {Abstreiter}}, \ and\ \bibinfo {author} {\bibfnamefont {J.~J.}\ \bibnamefont
  {Finley}},\ }\bibfield  {title} {\enquote {\bibinfo {title}
  {Observation and explanation of strong electrically tunable exciton $g$ factors in composition engineered In(Ga)As quantum dots},}\ }\href {\doibase 10.1103/PhysRevB.83.161303} {\bibfield
  {journal} {\bibinfo  {journal} {Phys. Rev. B}\ }\textbf {\bibinfo {volume}
  {83}},\ \bibinfo {pages} {161303} (\bibinfo {year} {2011})}\BibitemShut
  {NoStop}%
\bibitem [{\citenamefont {Beham}\ \emph {et~al.}(2001)\citenamefont {Beham},
  \citenamefont {Zrenner}, \citenamefont {Findeis}, \citenamefont {Bichler},\
  and\ \citenamefont {Abstreiter}}]{beham2001nonlinear}%
  \BibitemOpen
  \bibfield  {author} {\bibinfo {author} {\bibfnamefont {E.}~\bibnamefont
  {Beham}}, \bibinfo {author} {\bibfnamefont {A.}~\bibnamefont {Zrenner}},
  \bibinfo {author} {\bibfnamefont {F.}~\bibnamefont {Findeis}}, \bibinfo
  {author} {\bibfnamefont {M.}~\bibnamefont {Bichler}}, \ and\ \bibinfo
  {author} {\bibfnamefont {G.}~\bibnamefont {Abstreiter}},\ }\bibfield
  {title} {\enquote {\bibinfo {title} {Nonlinear ground-state absorption
  observed in a single quantum dot},}\ }\href {\doibase
  10.1063/1.1411987} {\bibfield  {journal} {\bibinfo  {journal} {Appl. Phys.
  Lett}\ }\textbf {\bibinfo {volume} {79}},\ \bibinfo {pages} {2808} (\bibinfo
  {year} {2001})}\BibitemShut {NoStop}%
\bibitem [{\citenamefont {Mar}\ \emph {et~al.}(2011{\natexlab{a}})\citenamefont
  {Mar}, \citenamefont {Xu}, \citenamefont {Baumberg}, \citenamefont {Irvine},
  \citenamefont {Stanley},\ and\ \citenamefont {Williams}}]{mar2011voltage}%
  \BibitemOpen
  \bibfield  {author} {\bibinfo {author} {\bibfnamefont {J.~D.}\ \bibnamefont
  {Mar}}, \bibinfo {author} {\bibfnamefont {X.~L.}\ \bibnamefont {Xu}},
  \bibinfo {author} {\bibfnamefont {J.~J.}\ \bibnamefont {Baumberg}}, \bibinfo
  {author} {\bibfnamefont {A.~C.}\ \bibnamefont {Irvine}}, \bibinfo {author}
  {\bibfnamefont {C.}~\bibnamefont {Stanley}}, \ and\ \bibinfo {author}
  {\bibfnamefont {D.~A.}\ \bibnamefont {Williams}},\ }\bibfield  {title}
  {\enquote {\bibinfo {title} {Voltage-controlled electron tunneling from a
  single self-assembled quantum dot embedded in a
  two-dimensional-electron-gas-based photovoltaic cell},}\ }\href {\doibase
  10.1063/1.3633216} {\bibfield  {journal} {\bibinfo  {journal} {J. Appl.
  Phys.}\ }\textbf {\bibinfo {volume} {110}},\ \bibinfo {pages} {053110}
  (\bibinfo {year} {2011}{\natexlab{a}})}\BibitemShut {NoStop}%
\bibitem [{\citenamefont {Mar}\ \emph {et~al.}(2011{\natexlab{b}})\citenamefont
  {Mar}, \citenamefont {Xu}, \citenamefont {Baumberg}, \citenamefont {Irvine},
  \citenamefont {Stanley},\ and\ \citenamefont
  {Williams}}]{mar2011electrically}%
  \BibitemOpen
  \bibfield  {author} {\bibinfo {author} {\bibfnamefont {J.~D.}\ \bibnamefont
  {Mar}}, \bibinfo {author} {\bibfnamefont {X.~L.}\ \bibnamefont {Xu}},
  \bibinfo {author} {\bibfnamefont {J.~J.}\ \bibnamefont {Baumberg}}, \bibinfo
  {author} {\bibfnamefont {A.~C.}\ \bibnamefont {Irvine}}, \bibinfo {author}
  {\bibfnamefont {C.}~\bibnamefont {Stanley}}, \ and\ \bibinfo {author}
  {\bibfnamefont {D.~A.}\ \bibnamefont {Williams}},\ }\bibfield  {title}
  {\enquote {\bibinfo {title} {Electrically tunable hole tunnelling from a
  single self-assembled quantum dot embedded in an n-i-Schottky photovoltaic
  cell},}\ }\href {\doibase
  10.1063/1.3614418} {\bibfield  {journal} {\bibinfo  {journal} {Appl. Phys.
  Lett}\ }\textbf {\bibinfo {volume} {99}},\ \bibinfo {pages} {031102}
  (\bibinfo {year} {2011}{\natexlab{b}})}\BibitemShut {NoStop}%
\bibitem [{\citenamefont {Patan\`e}\ \emph {et~al.}(2002)\citenamefont
  {Patan\`e}, \citenamefont {Hill}, \citenamefont {Eaves}, \citenamefont
  {Main}, \citenamefont {Henini}, \citenamefont {Zambrano}, \citenamefont
  {Levin}, \citenamefont {Mori}, \citenamefont {Hamaguchi}, \citenamefont
  {Dubrovskii}, \citenamefont {Vdovin}, \citenamefont {Austing}, \citenamefont
  {Tarucha},\ and\ \citenamefont {Hill}}]{patan2002probing}%
  \BibitemOpen
  \bibfield  {author} {\bibinfo {author} {\bibfnamefont {A.}~\bibnamefont
  {Patan\`e}}, \bibinfo {author} {\bibfnamefont {R.~J.~A.}\ \bibnamefont
  {Hill}}, \bibinfo {author} {\bibfnamefont {L.}~\bibnamefont {Eaves}},
  \bibinfo {author} {\bibfnamefont {P.~C.}\ \bibnamefont {Main}}, \bibinfo
  {author} {\bibfnamefont {M.}~\bibnamefont {Henini}}, \bibinfo {author}
  {\bibfnamefont {M.~L.}\ \bibnamefont {Zambrano}}, \bibinfo {author}
  {\bibfnamefont {A.}~\bibnamefont {Levin}}, \bibinfo {author} {\bibfnamefont
  {N.}~\bibnamefont {Mori}}, \bibinfo {author} {\bibfnamefont {C.}~\bibnamefont
  {Hamaguchi}}, \bibinfo {author} {\bibfnamefont {Y.~V.}\ \bibnamefont
  {Dubrovskii}}, \bibinfo {author} {\bibfnamefont {E.~E.}\ \bibnamefont
  {Vdovin}}, \bibinfo {author} {\bibfnamefont {D.~G.}\ \bibnamefont {Austing}},
  \bibinfo {author} {\bibfnamefont {S.}~\bibnamefont {Tarucha}}, \ and\
  \bibinfo {author} {\bibfnamefont {G.}~\bibnamefont {Hill}},\ }\bibfield
  {title} {\enquote {\bibinfo {title} {Probing the quantum states of
  self-assembled InAs dots by magnetotunneling spectroscopy},}\ }\href {\doibase
  10.1103/PhysRevB.65.165308} {\bibfield  {journal} {\bibinfo  {journal} {Phys.
  Rev. B}\ }\textbf {\bibinfo {volume} {65}},\ \bibinfo {pages} {165308}
  (\bibinfo {year} {2002})}\BibitemShut {NoStop}%
\bibitem [{\citenamefont {Godden}\ \emph
  {et~al.}(2012{\natexlab{b}})\citenamefont {Godden}, \citenamefont {Quilter},
  \citenamefont {Ramsay}, \citenamefont {Wu}, \citenamefont {Brereton},
  \citenamefont {Luxmoore}, \citenamefont {Puebla}, \citenamefont {Fox},\ and\
  \citenamefont {Skolnick}}]{godden2012fast}%
  \BibitemOpen
  \bibfield  {author} {\bibinfo {author} {\bibfnamefont {T.~M.}\ \bibnamefont
  {Godden}}, \bibinfo {author} {\bibfnamefont {J.~H.}\ \bibnamefont {Quilter}},
  \bibinfo {author} {\bibfnamefont {A.~J.}\ \bibnamefont {Ramsay}}, \bibinfo
  {author} {\bibfnamefont {Y.}~\bibnamefont {Wu}}, \bibinfo {author}
  {\bibfnamefont {P.}~\bibnamefont {Brereton}}, \bibinfo {author}
  {\bibfnamefont {I.~J.}\ \bibnamefont {Luxmoore}}, \bibinfo {author}
  {\bibfnamefont {J.}~\bibnamefont {Puebla}}, \bibinfo {author} {\bibfnamefont
  {A.~M.}\ \bibnamefont {Fox}}, \ and\ \bibinfo {author} {\bibfnamefont
  {M.~S.}\ \bibnamefont {Skolnick}},\ }\bibfield  {title} {\enquote {\bibinfo
  {title} {Fast preparation of a single-hole spin in an InAs/GaAs quantum dot
  in a Voigt-geometry magnetic field},}\ }\href {\doibase
  10.1103/PhysRevB.85.155310} {\bibfield  {journal} {\bibinfo  {journal} {Phys.
  Rev. B}\ }\textbf {\bibinfo {volume} {85}},\ \bibinfo {pages} {155310}
  (\bibinfo {year} {2012}{\natexlab{b}})}\BibitemShut {NoStop}%
\bibitem [{\citenamefont {Stevenson}\ \emph {et~al.}(2006)\citenamefont
  {Stevenson}, \citenamefont {Young}, \citenamefont {See}, \citenamefont
  {Gevaux}, \citenamefont {Cooper}, \citenamefont {Atkinson}, \citenamefont
  {Farrer}, \citenamefont {Ritchie},\ and\ \citenamefont
  {Shields}}]{stevenson2006magnetic}%
  \BibitemOpen
  \bibfield  {author} {\bibinfo {author} {\bibfnamefont {R.~M.}\ \bibnamefont
  {Stevenson}}, \bibinfo {author} {\bibfnamefont {R.~J.}\ \bibnamefont
  {Young}}, \bibinfo {author} {\bibfnamefont {P.}~\bibnamefont {See}}, \bibinfo
  {author} {\bibfnamefont {D.~G.}\ \bibnamefont {Gevaux}}, \bibinfo {author}
  {\bibfnamefont {K.}~\bibnamefont {Cooper}}, \bibinfo {author} {\bibfnamefont
  {P.}~\bibnamefont {Atkinson}}, \bibinfo {author} {\bibfnamefont
  {I.}~\bibnamefont {Farrer}}, \bibinfo {author} {\bibfnamefont {D.~A.}\
  \bibnamefont {Ritchie}}, \ and\ \bibinfo {author} {\bibfnamefont {A.~J.}\
  \bibnamefont {Shields}},\ }\bibfield  {title} {\enquote {\bibinfo {title}
  {Magnetic-field-induced reduction of the exciton polarization splitting in
  InAs quantum dots},}\ }\href {\doibase 10.1103/PhysRevB.73.033306}
  {\bibfield  {journal} {\bibinfo  {journal} {Phys. Rev. B}\ }\textbf {\bibinfo
  {volume} {73}},\ \bibinfo {pages} {033306} (\bibinfo {year}
  {2006})}\BibitemShut {NoStop}%
\bibitem [{\citenamefont {Stevenson}\ \emph {et~al.}(2004)\citenamefont
  {Stevenson}, \citenamefont {Young}, \citenamefont {See}, \citenamefont
  {Farrer}, \citenamefont {Ritchie},\ and\ \citenamefont
  {Shields}}]{stevenson2004time}%
  \BibitemOpen
  \bibfield  {author} {\bibinfo {author} {\bibfnamefont {R.}~\bibnamefont
  {Stevenson}}, \bibinfo {author} {\bibfnamefont {R.}~\bibnamefont {Young}},
  \bibinfo {author} {\bibfnamefont {P.}~\bibnamefont {See}}, \bibinfo {author}
  {\bibfnamefont {I.}~\bibnamefont {Farrer}}, \bibinfo {author} {\bibfnamefont
  {D.}~\bibnamefont {Ritchie}}, \ and\ \bibinfo {author} {\bibfnamefont
  {A.}~\bibnamefont {Shields}},\ }\bibfield  {title} {\enquote {\bibinfo
  {title} {Time-resolved studies of single quantum dots in magnetic fields},}\
  }\href {\doibase
  http://dx.doi.org/10.1016/j.physe.2003.11.033} {\bibfield  {journal}
  {\bibinfo  {journal} {Physica E}\ }\textbf {\bibinfo {volume} {21}},\
  \bibinfo {pages} {381} (\bibinfo {year} {2004})}\BibitemShut {NoStop}%
\bibitem [{\citenamefont {Zieli\ifmmode~\acute{n}\else \'{n}\fi{}ski}\ \emph
  {et~al.}(2015)\citenamefont {Zieli\ifmmode~\acute{n}\else \'{n}\fi{}ski},
  \citenamefont {Don},\ and\ \citenamefont {Gershoni}}]{Zielinski2015}%
  \BibitemOpen
  \bibfield  {author} {\bibinfo {author} {\bibfnamefont {M.}~\bibnamefont
  {Zieli\ifmmode~\acute{n}\else \'{n}\fi{}ski}}, \bibinfo {author}
  {\bibfnamefont {Y.}~\bibnamefont {Don}}, \ and\ \bibinfo {author}
  {\bibfnamefont {D.}~\bibnamefont {Gershoni}},\ }\bibfield  {title} {\enquote
  {\bibinfo {title} {Atomistic theory of dark excitons in self-assembled
  quantum dots of reduced symmetry},}\ }\href {\doibase
  10.1103/PhysRevB.91.085403} {\bibfield  {journal} {\bibinfo  {journal} {Phys.
  Rev. B}\ }\textbf {\bibinfo {volume} {91}},\ \bibinfo {pages} {085403}
  (\bibinfo {year} {2015})}\BibitemShut {NoStop}%
\bibitem [{\citenamefont {Stufler}\ \emph {et~al.}(2004)\citenamefont
  {Stufler}, \citenamefont {Ester}, \citenamefont {Zrenner},\ and\
  \citenamefont {Bichler}}]{stufler2004power}%
  \BibitemOpen
  \bibfield  {author} {\bibinfo {author} {\bibfnamefont {S.}~\bibnamefont
  {Stufler}}, \bibinfo {author} {\bibfnamefont {P.}~\bibnamefont {Ester}},
  \bibinfo {author} {\bibfnamefont {A.}~\bibnamefont {Zrenner}}, \ and\
  \bibinfo {author} {\bibfnamefont {M.}~\bibnamefont {Bichler}},\ }\bibfield  {title} {\enquote {\bibinfo {title} {Power broadening of the
  exciton linewidth in a single InGaAs/GaAs quantum dot},}\ }\href
  {\doibase 10.1063/1.1815373} {\bibfield  {journal} {\bibinfo  {journal}
  {Appl. Phys. Lett}\ }\textbf {\bibinfo {volume} {85}},\ \bibinfo {pages}
  {4202} (\bibinfo {year} {2004})}\BibitemShut {NoStop}%
\bibitem [{\citenamefont {Stufler}\ \emph {et~al.}(2005)\citenamefont
  {Stufler}, \citenamefont {Ester}, \citenamefont {Zrenner},\ and\
  \citenamefont {Bichler}}]{stufler2005quantum}%
  \BibitemOpen
  \bibfield  {author} {\bibinfo {author} {\bibfnamefont {S.}~\bibnamefont
  {Stufler}}, \bibinfo {author} {\bibfnamefont {P.}~\bibnamefont {Ester}},
  \bibinfo {author} {\bibfnamefont {A.}~\bibnamefont {Zrenner}}, \ and\
  \bibinfo {author} {\bibfnamefont {M.}~\bibnamefont {Bichler}},\ }\bibfield
  {title} {\enquote {\bibinfo {title} {Quantum optical properties of a single
  ${\mathrm{In}}_{x}{\mathrm{Ga}}_{1\ensuremath{-}x}\mathrm{As}\text{\ensuremath{-}}\mathrm{Ga}\mathrm{As}$
  quantum dot two-level system},}\ }\href
  {\doibase 10.1103/PhysRevB.72.121301} {\bibfield  {journal} {\bibinfo
  {journal} {Phys. Rev. B}\ }\textbf {\bibinfo {volume} {72}},\ \bibinfo
  {pages} {121301} (\bibinfo {year} {2005})}\BibitemShut {NoStop}%
\end{thebibliography}
%

\end{document}